\documentclass[twocolumn]{aastex62}
\usepackage{lineno}

\newcommand{\eqb}{\begin{eqnarray}}
\newcommand{\eqe}{\end{eqnarray}}
\newcommand{\beq}{\begin{equation}}
\newcommand{\eeq}{\end{equation}} 
\newcommand{\tvar}{t_{\rm v}}
\newcommand{\dop}{\mathcal{D}}
\newcommand{\pg}{p\pi}
\newcommand{\gp}{\gamma^\prime_p}
\shorttitle{Energy crisis in blazars}
\shortauthors{Liodakis \& Petropoulou}
\begin{document}

\title{Proton Synchrotron $\gamma$-rays and the Energy Crisis in Blazars}
%Minimum jet power and the energy crisis in blazars: the proton synchrotron case

\correspondingauthor{I. Liodakis}
\email{ilioda@stanford.edu}

\correspondingauthor{M. Petropoulou}
\email{m.petropoulou@astro.princeton.edu}

\author{Ioannis Liodakis}
\affil{KIPAC, Stanford University, 452 Lomita Mall, Stanford, CA 94305, USA}

\author{Maria Petropoulou}
\affil{Department of Astrophysical Sciences, Princeton University, Princeton, NJ 08544, USA}

\begin{abstract}
The origin of high-energy emission in blazars jets (i.e., leptonic versus hadronic) has been a long-standing matter of debate. Here, we focus on one variant of hadronic models where proton synchrotron radiation accounts for the observed steady $\gamma$-ray blazar emission. Using analytical methods, we derive the minimum jet power ($P_{j,\min}$) for the largest blazar sample analyzed to date (145 sources), taking into account uncertainties of observables and jet's physical parameters. We compare $P_{j,\min}$  against three characteristic energy estimators for accreting systems, i.e., the Eddington luminosity, the accretion disk luminosity, and the power of the Blandford-Znajek process, and find that $P_{j,\min}$ is about 2 orders of magnitude higher than all energetic estimators for the majority of our sample. The derived magnetic field strengths in the emission region require either large  amplification of the jet's magnetic field (factor of 30) or place the $\gamma$-ray production site at sub-pc scales. The expected neutrino emission peaks at $\sim 0.1-10$~EeV, with typical peak neutrino fluxes $\sim 10^{-4}$ times lower than the peak $\gamma$-ray fluxes. We conclude that if relativistic hadrons are present in blazar jets, they can only produce a radiatively subdominant component of the overall spectral energy distribution of the blazar's steady emission.
\end{abstract}
\keywords{black hole physics ---  radiation mechanisms: non-thermal --- galaxies: active --- galaxies: jets}

\section{Introduction}\label{introduc}
High-energy emission of blazars --  active galactic nuclei with relativistic jets closely aligned to our line of sight, powered by accretion onto a supermassive black hole (BH) -- has been a matter of vibrant debate since their first  detection in $\gamma$-rays \citep[for a review, see][]{Blandford2018}.

Historically, $\gamma$-ray emission has been attributed to two broad classes of models that are distinguished mainly by the species of radiating particles. Leptonic models invoke inverse Compton scattering of low-energy photons
by relativistic electrons \citep[e.g.,][]{Marscher1985,Dermer1992}.
Hadronic models involve a variety of mechanisms that are directly or indirectly  related to relativistic hadrons, such as proton synchrotron radiation \citep[e.g.,][]{Aharonian2000}, or synchrotron and Compton processes of secondary electrons and positrons produced in photo-hadronic interactions \citep[e.g.,][]{Mannheim1993}.

Unveiling the dominant process for blazar's $\gamma$-ray emission has been the subject of numerous studies. This is not surprising, since by constraining the dominant high-energy processes in blazars we can probe the jet's physical conditions (which are hidden to direct observation) and help answer long-standing questions regarding launching and mass-loading of jets \citep{Blandford2018}.

The most common methods to probe the origin of $\gamma$-rays are spectral energy distribution (SED) modeling of broadband emission \citep[e.g.,][]{Boettcher2013,Ghisellini2015, Petropoulou2015} and searches for correlated variability between low-energy radiation (e.g., radio and optical) and $\gamma$-rays \citep[e.g.,][]{Max-Moerbeck2014,Liodakis2018-II,Liodakis2019}. While past studies have favored leptonic models, they have not been always conclusive. The most recent possible association of high energy neutrinos with blazar TXS~0506+056 \citep{IceCube2018, Aartsen2018b} would also suggest that the usually disfavored hadronic component should be present. Interestingly, SED modeling of the first likely multi-messenger event point to leptonic processes dominating the $\gamma$-ray emission \citep[e.g.,][]{Gao2019,Zhang2020}, although the jet's energetics are still governed by relativistic hadrons \citep{Keivani2018,Petropoulou2020}.

Indeed, one of the major criticisms of hadronic models for blazar emission relates to their energetic requirements. The inefficiency of hadronic processes was pointed out using generic arguments by \citet{Sikora2009,Sikora2011}, and later discussed on a source-to-source basis using SED modeling of steady emission or $\gamma$-ray flares \citep[e.g.,][]{Cerruti2015, Petrodimi2015, Petropoulou2017}.
Recently, \cite{Zdziarski2015} explored the energetic requirements of the proton synchrotron (PS) model for a limited sample \citep[12 sources from][]{Boettcher2013}, and concluded that the estimated minimum jet powers are not compatible with the inferred accretion power and Eddington luminosity. 

In light of recent results, we revisit the jet-power analysis of $\gamma$-ray blazars in the PS model by following the analytical approach of \citet{Petropoulou2016-II} and extending our calculations to the largest sample to date (145 sources).
 
\section{Sample}\label{sec:sample}

Our sample consists of sources with synchrotron peak frequency and luminosity from the 4th {\it Fermi} AGN catalog (4LAC, \citealp{4LAC2019}),  Doppler factors from  \cite{Liodakis2018}, and apparent velocities ($\beta_{app}$) from the MOJAVE survey \cite{Lister2016}. We use the SED builder tool\footnote{\url{https://tools.ssdc.asi.it/SED/}} of the Space Science Data Center (SSDC) to fit a third degree polynomial in log-log space to the archival data (similar to the analysis of the 4LAC for the synchrotron spectrum) to estimate the peak frequency and luminosity of the high-energy component. We have removed sources where the data are insufficient to confidently determine the parameters of the high-energy component; these sources either lacked X-ray observations that constrain the low-energy part of the high-energy component or  $\gamma$-ray observations were not available through SSDC. Our final sample consists of 145 sources.

\section{Methods}\label{sec:method}
The absolute jet power for a two-sided jet can be written as
$P_j = 2\pi r^{\prime 2} \beta \Gamma^2 c \sum_{i=B,e,p} \!\!\left(u_i^\prime +p^\prime_i\right)+ P_{j}^{r} +P_{j}^{c}$,
where $r^\prime$ is the radius of the emitting region\footnote{Quantities measured in the jet's co-moving frame are noted with primes.},  $\Gamma=(1-\beta^2)^{-1/2}$ is the jet's bulk Lorentz factor, $u_i^\prime$ is the energy density of relativistic particles/magnetic fields and $p^\prime_i=u_i^\prime/3$, $P_j^{r}$ is the absolute photon luminosity, and $P_{j}^{c}$ is the contribution of cold protons to the total jet power \citep{2014MNRAS.445.1321Z, Petropoulou2016-II}.  Henceforth, we drop the latter term from our analysis, as it is negligible compared to the others in the PS scenario. 

Following \cite{Petropoulou2016-II} (henceforth, PD16), we assume monoenergetic particle distributions for both relativistic electrons and protons (i.e., $N_i(\gamma_i^\prime)=N_{i,0}\delta(\gamma_i^\prime-\bar{\gamma}^\prime_i)$, $i=e,p$). This choice is equivalent to the assumption of power-law particle energy spectra with slopes $p<2$, while consideration of steep power laws ($p>2$) would only increase our minimum power estimates. Same as in PD16, we assume that the proton radiative efficiency is $\simeq 1$ (lower efficiency would only increase the energetic requirements). We can re-write $P_j$ as a function of the  emitting region's Doppler factor $\mathcal{D}=\left[\Gamma(1-\beta \cos\theta)\right]^{-1}$ (here, $\theta$ is the observer's angle) and co-moving magnetic field strength $x \equiv B^\prime/B_{\rm cr}$ (in units of $B_{\rm cr}=4.4\times10^{13}$~G),
\eqb
\psi^{-2} P_j &=& A_B(\tvar, z) x^2 \mathcal{D}^{4} + \nonumber \\ 
& & A_e(L_{l}, \varepsilon_{l}, \tvar, z) x^{-3/2}\mathcal{D}^{-5/2}\left(1+ \frac{2A_e}{L_l} x^{-3/2}\mathcal{D}^{-1/2} \right) + \nonumber \\ 
& & A_p(L_{h}, \varepsilon_{h}, \tvar, z) x^{-3}\mathcal{D}^{-4}\left(1+ \frac{2A_p}{L_h} x^{-3}\mathcal{D}^{-2} \right) + \nonumber \\ 
& & A_r(L_h/L_l) \mathcal{D}^{-2} L_l.
\label{eq:Pjet}
\eqe 
Here, $\psi = 1 + (\Gamma \theta)^2 \approx 2\Gamma/\dop$, and $A_i(\cdots)$ with $i=B,e,p,r$ are functions of source parameters\footnote{For the full expressions, see PD16. The correspondence  in notation between the two papers is: $A_B \rightarrow \mathcal{A}$, $A_e\rightarrow\mathcal{B}$, $A_p\rightarrow\mathcal{C}$, and $A_r\rightarrow\mathcal{E}$. The $z$ dependence was not included in PD16.}: redshift $z$, typical variability timescale $\tvar$, peak luminosities of the low- and high-energy SED humps, $L_{l}$ and $L_{h}$, and the respective peak photon energies $\varepsilon_{l}$ and $\varepsilon_{h}$ (both in units of $m_e c^2$). Knowledge of the SED parameters, relativistic boosting effects, and variability timescale of a source allows us to estimate the minimum jet power ($P_{j,\min}$) with respect to the unknown variable $B^\prime$. 

For each source we derive $P_{j,\min}$ and the corresponding magnetic field strength $B^\prime$ for $10^4$ combinations of random values for $\varepsilon_{l,h}$, $L_{l,h}$, $\tvar$, $\dop$, and $\beta_{app}$ drawn from Gaussian distributions  with mean $\mu$ and standard deviation $\sigma$. 
For $\tvar$ we choose $\mu=10^5$~s and $\sigma=3\times10^4$~s, which translates to a range of minutes to $>2$ days (e.g., \citealp{Meyer2019}). We assume a $\sigma$ of 0.5~dex for the luminosities and for the peak frequencies 0.3~dex \citep{Lister2015}. For  $\dop$ and $\beta_{app}$ needed to estimate $\Gamma$ and $\theta$, we use the values and their uncertainties listed in \cite{Liodakis2018}.

To assess our results, we compare the derived $P_{j,\min}$ to three characteristic ``energy estimators'' of an accreting BH system: (i) the Eddington luminosity $L_{\rm Edd}$, (ii) the accretion disk luminosity $L_d$, and (iii) the power of the Blandford-Znajek (BZ) process $P_{\rm BZ}$ \citep{Blandford1977}. We estimate the BH masses for 82 blazars in our sample (needed for computing $L_{\rm Edd}$ and  $P_{\rm BZ}$) using the H$\beta$, MgII, and CIV full width at half maximum and line luminosities from \cite{Shaw2012} and \cite{Torrealba2012}, together with the scaling relations from \citet[][Equation 5, Table 2]{Shaw2012}.  We complement our sample with 13 mass estimates from \cite{Woo2002,Wang2004,Liu2006} that use the same lines. For the remaining sources we use the BL Lac and Flat Spectrum Radio Quasar (FSRQ) population median and standard deviations derived from the BH estimates in this work.  The accretion disk luminosity (71 sources) is estimated using the line luminosities of the H$\beta$, MgII, and OIII lines and the scaling relations from \citet[Equations 9, 10, 11]{Zamaninasab2014}.  When multiple estimates for either the BH mass or $L_d$ are available we use the median of the estimates and for its uncertainty we quote the standard deviation of the estimates or the average uncertainty (whichever is greater).

To estimate the power of the BZ process, we first estimate the jet's co-moving magnetic field strength at 1~parsec (pc), $B^\prime_{\rm 1 pc}$, using the core-shift measurements  and equations (2) and (3) from \cite{Zamaninasab2014}, with the correct redshift terms \citep{Lobanov1998,Zdziarski2015-II}. We then derive the poloidal magnetic flux that threads the pc-scale jet, $\Phi_{\rm jet}$, using the jet apparent opening angles from \cite{Pushkarev2009} and equation (1) from  \cite{Zamaninasab2014}. This quantity is a proxy of the poloidal magnetic flux threading the BH ($\Phi_{\rm BH}$) under the flux-freezing assumption.
For the BH spin $a$, we consider three cases: all BHs are maximally spinning;  $a$ follows a uniform distribution from 0 to 1; $a$ follows a Beta distribution\footnote{This is parametric probability distribution defined between 0 and 1 as $P(x)=(1-x)^{\beta-1}x^{\alpha-1}/B(\alpha,\beta)$, where $B(\alpha,\beta)$ is a Beta function, and $\alpha, \beta$ are shape parameters related to $\mu, \sigma$ as $\mu=\alpha/(\alpha+\beta)$ and $\sigma=(\alpha\beta/(\alpha+\beta+1)(\alpha+\beta)^2)^{1/2}$.} with $\mu=0.937$, $\sigma=0.074$ for BL Lacs and $\mu=0.742$, $\sigma=0.163$ for FSRQs \citep{Liodakis2018-III}. The BZ power is then estimated as $P_{\rm BZ}=\kappa\Omega_{\rm H}^2\Phi_{\rm BH}^2f(\Omega_{\rm H})/4{\pi}c$, where $\rm\kappa\approx0.05$ is a numerical constant whose value depends on the magnetic  field  geometry, $\Omega_{\rm H}=a c/2r_{\rm H}$ is the angular frequency of the BH horizon, $r_{\rm H}=r_ g(1+\sqrt{1-a})$ is the BH event horizon radius, $r_g=GM_{\bullet}/c^2$ is the gravitational radius, $c$ is the speed of light, $M_{\bullet}$ is the BH mass, and $f(\Omega_{\rm H})$ as
$f(\Omega_{\rm H})\approx1+1.38(\Omega_{\rm H} r_g/c)^2-9.2(\Omega_{\rm H} r_g/c)^4$ \citep{Tchekhovskoy2011}. All derived parameters are listed in Table~\ref{tab:par_sources}.

\section{Results}\label{sec:res}
\begin{figure}
\includegraphics[width=0.48\textwidth]{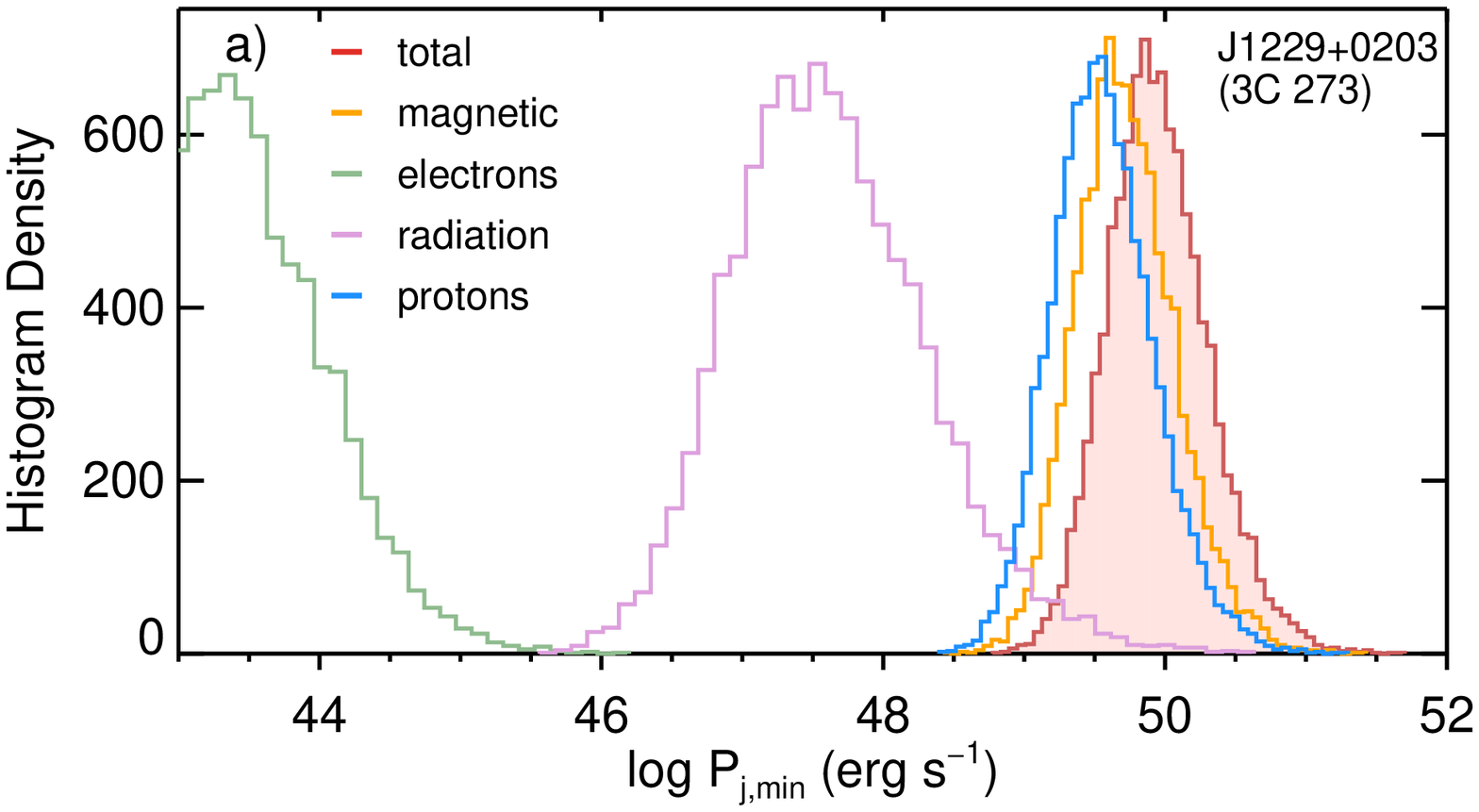}
\includegraphics[width=0.48\textwidth]{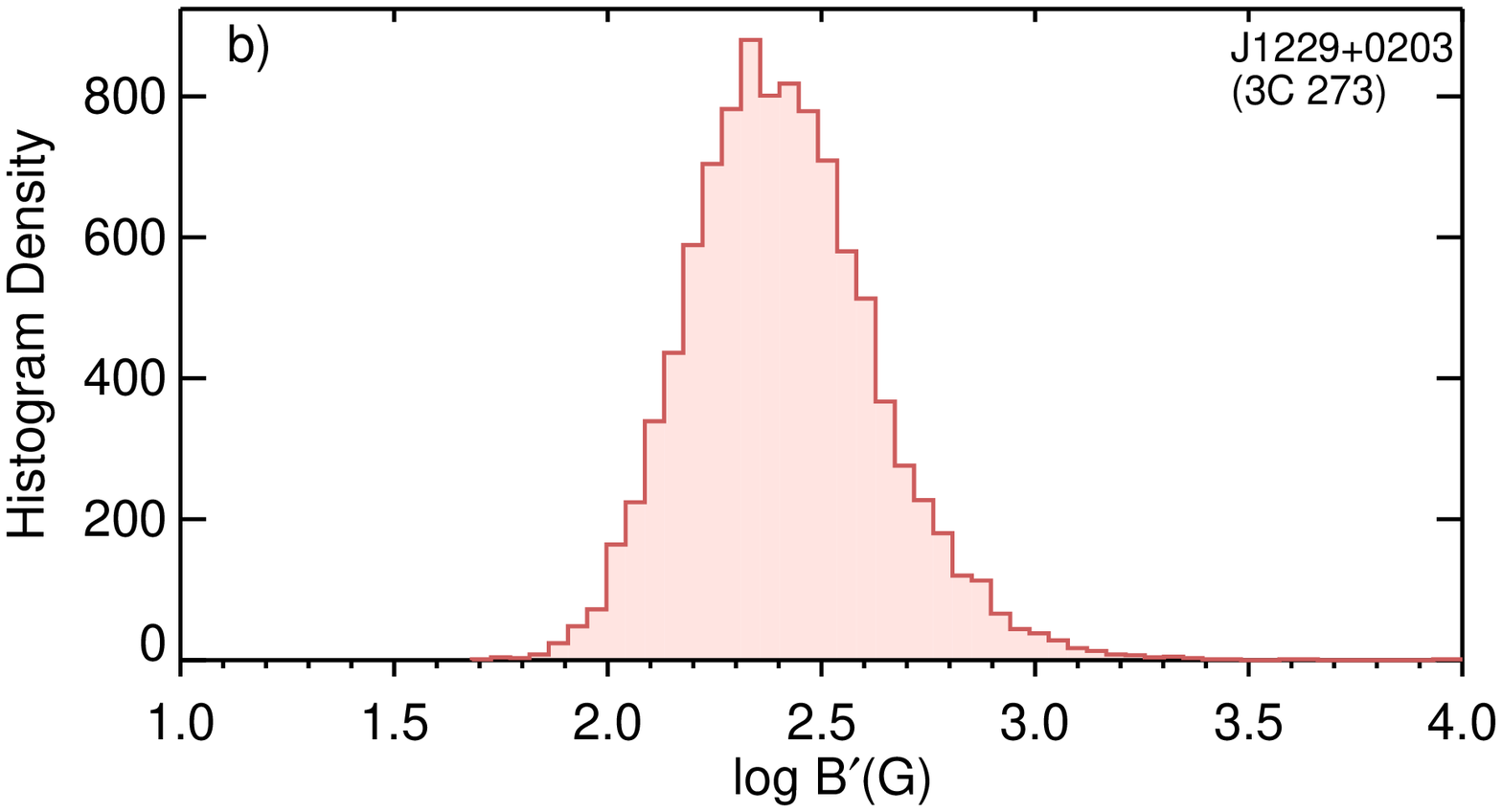}
\includegraphics[width=0.48\textwidth]{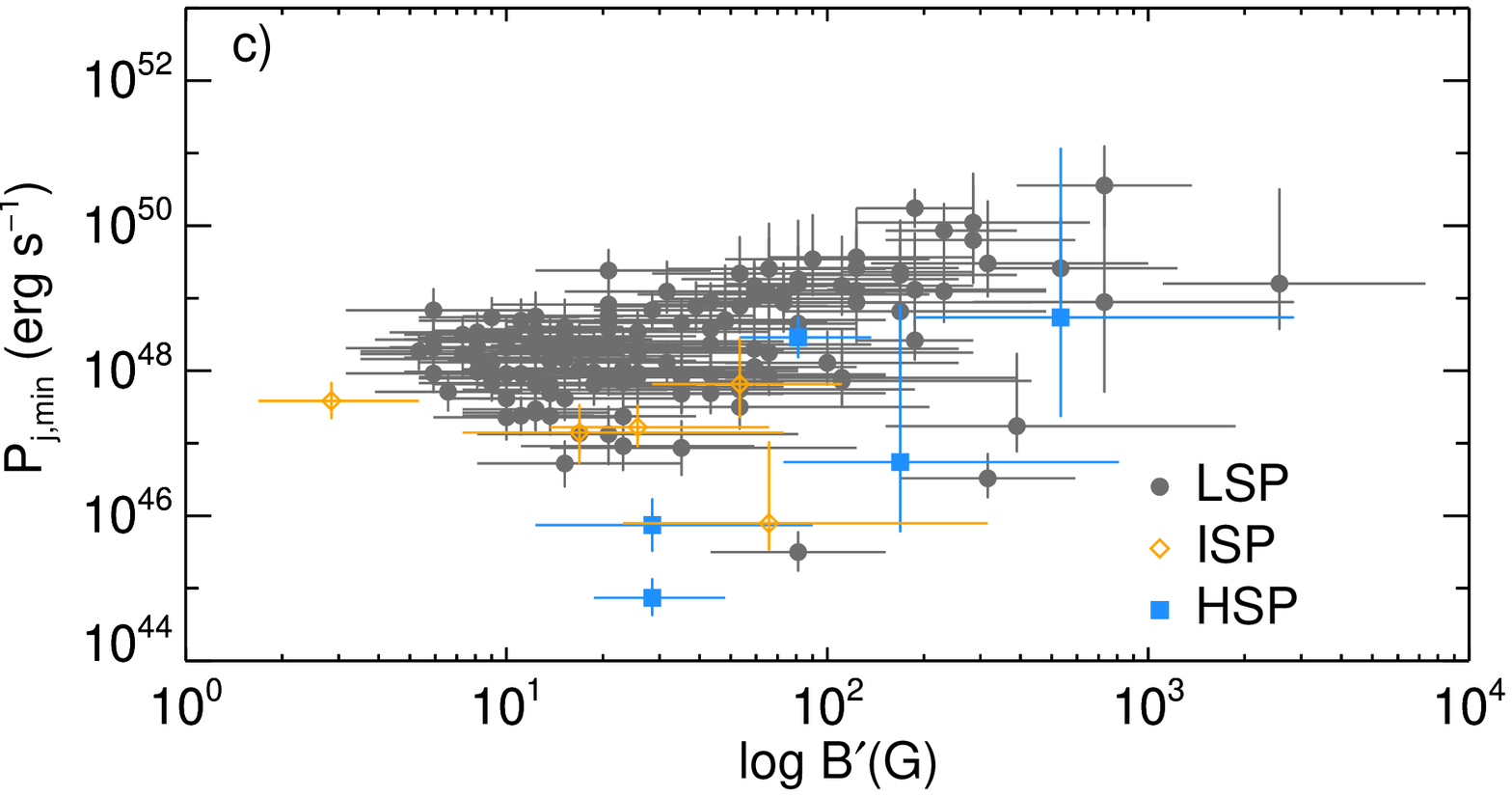}
\caption{{\bf Panel a}: Distribution of the minimum total jet power (filled histogram) and its different components (open colored histograms) for 3C~273. {\bf Panel b}: Distribution of co-moving magnetic field strength that minimizes the total jet power  for 3C~273. {\bf Panel c}:  Scatter plot of the median values of $P_{j,\min}$ and  $B^\prime$ for all sources from our sample, with error bars showing the 68\% uncertainty. Different symbols show blazar spectral types (see inset legend).}
\label{plt:hist_B_P}
\end{figure}

Panels a and b of Figure \ref{plt:hist_B_P}  show, respectively, the distributions of $P_{j,\min}$ and $B^\prime$ that minimizes the total jet power for 3C~273 . The magnetic field and relativistic proton components contribute the most to the total jet power, as expected in the PS scenario (PD16). Our analytical method yields $P_{j,\min}=8.6^{+11.7}_{-4.6}\times10^{49}$~erg s$^{-1}$ which is consistent with SED modeling results \citep{Petrodimi2015, Boettcher2013}. We also find comparable (within 90\% uncertainty) $P_{j,\min}$ values for 10 sources we have in common with \cite{Boettcher2013}. Panel c of Figure~\ref{plt:hist_B_P} shows the results for the whole sample. Different symbols are used to identify blazar classes according to their peak (rest-frame) synchrotron frequency: low-synchrotron peaked (LSP) sources ($\nu_s<10^{14}$~Hz), intermediate-synchrotron peaked (ISP) ($10^{14}<\nu_s<10^{15}$~Hz), and high-synchrotron peaked ($\nu_s>10^{15}$~Hz) sources \citep{Abdo2010}. The minimum jet power decreases on average as we move from LSP to HSP sources (PD16), while blazars with higher $P_{j,\min}$ tend to have stronger magnetic fields in the emission region.

\begin{figure}
\includegraphics[width=0.48\textwidth]{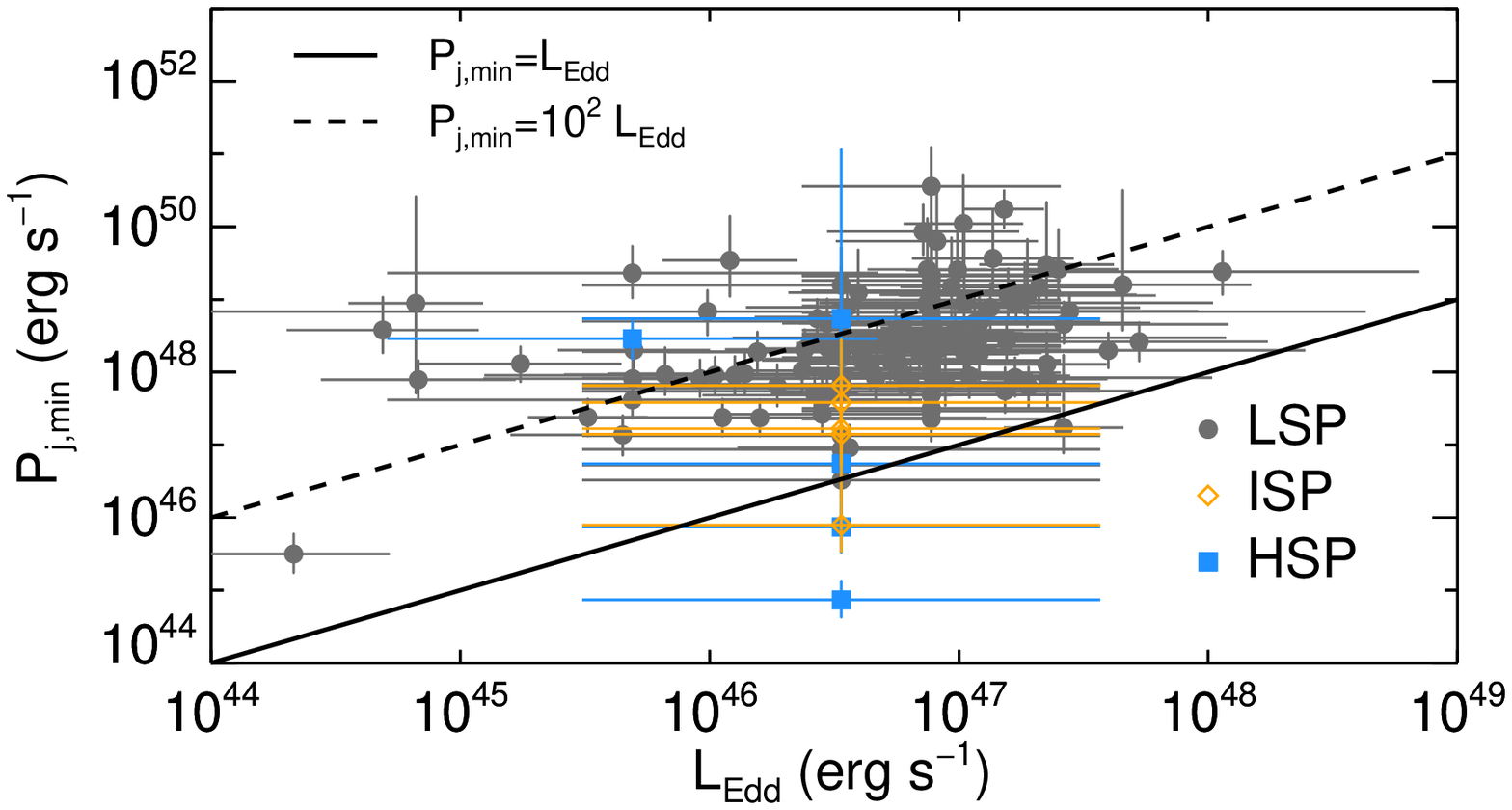}
\includegraphics[width=0.48\textwidth]{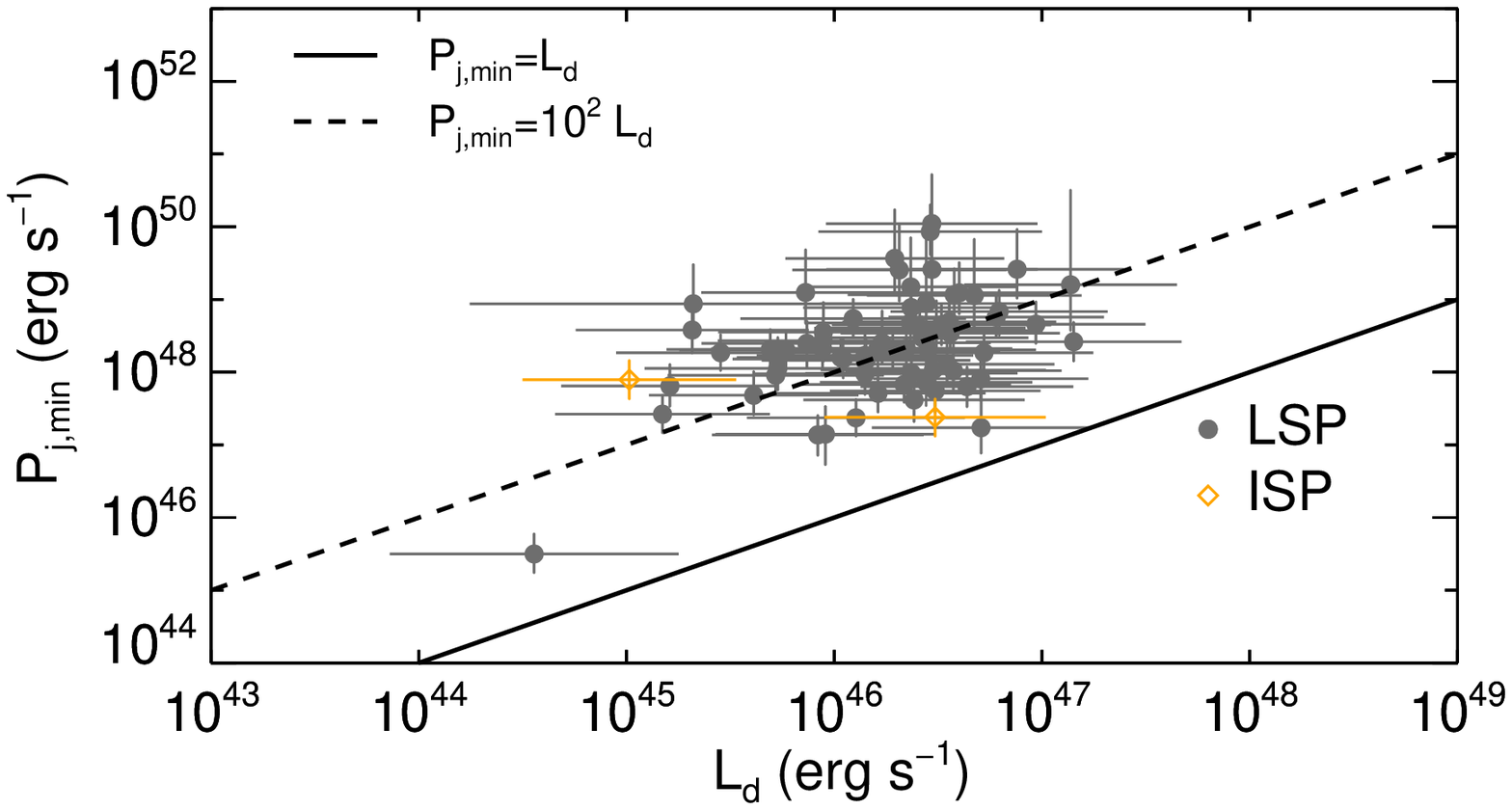}
 \caption{{\bf Top panel}: Minimum jet power versus Eddington luminosity. {\bf Bottom panel}: Minimum jet power versus accretion disk luminosity.
 }
\label{plt:P_v_L}
\end{figure}

\begin{figure}
\includegraphics[width=0.48\textwidth]{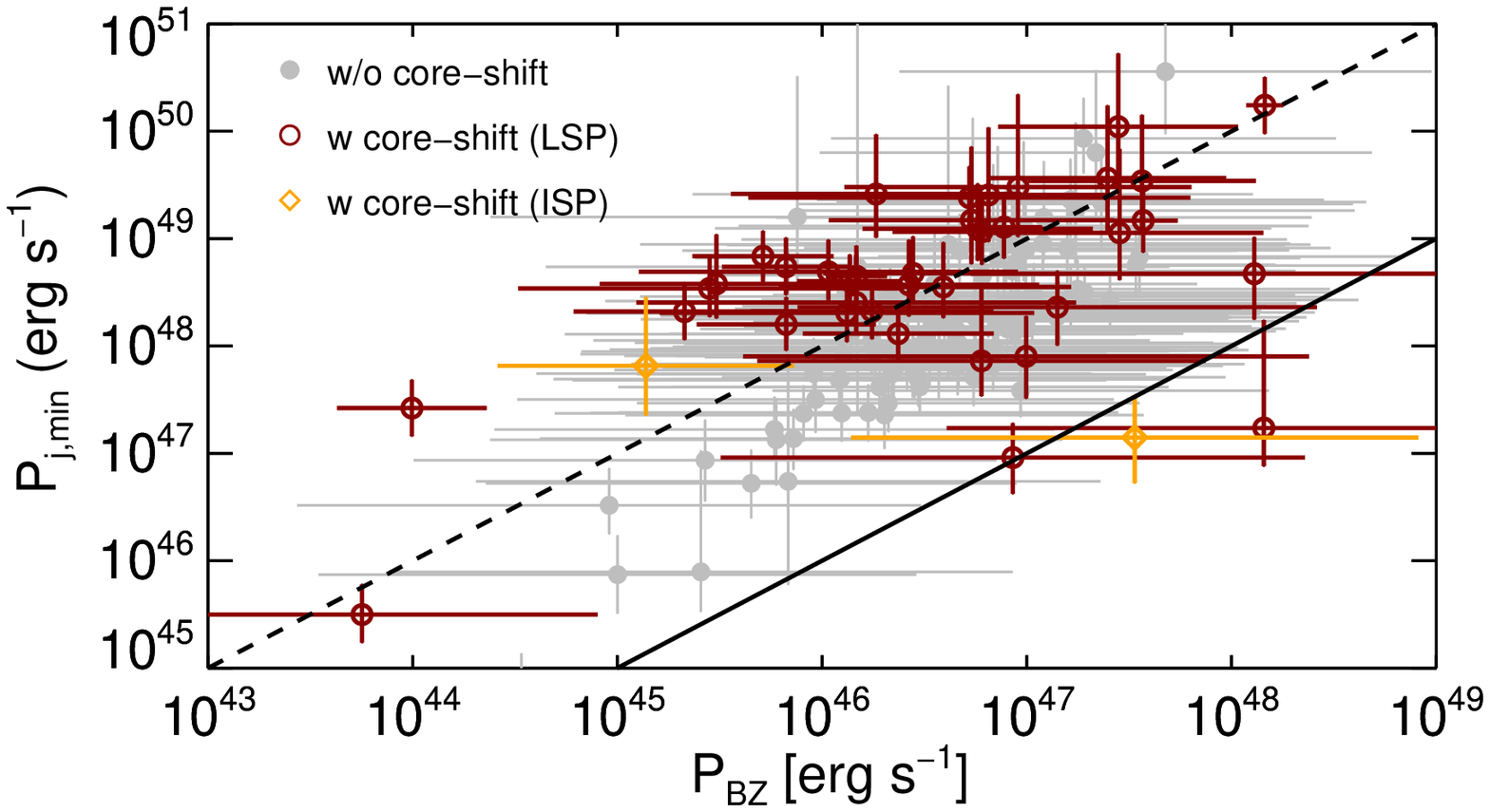}
\includegraphics[width=0.48\textwidth]{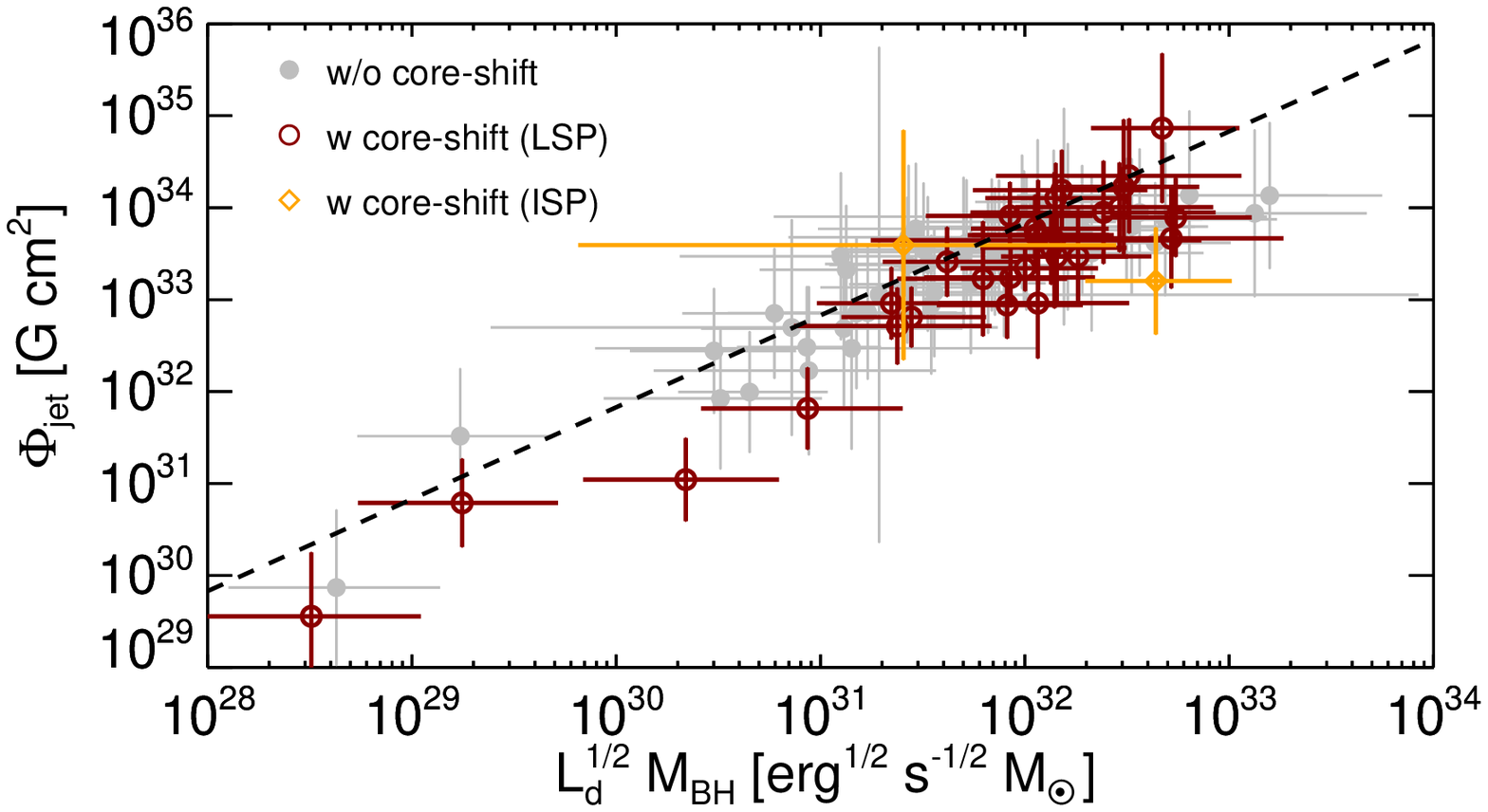}
 \caption{{\bf Top panel:} Minimum jet power versus the Blandford-Znajek power estimated for the optimistic case of maximally spinning black holes.  Solid (dashed) lines indicate the relation $P_{j,\min}=P_{\rm BZ}$ ($P_{j,\min}=10^2 P_{\rm BZ}$). {\bf Bottom panel:} Magnetic flux of the jet versus $L^{1/2}_{d} M_{\odot}$ for a sub-sample with $L_d$ measurements. The  prediction of a magnetically arrested disk overplotted (dashed line). In both panels, sources with and without core-shift measurements are plotted with open and filled symbols, respectively.}
\label{plt:P_v_P}
\end{figure}

Figure \ref{plt:P_v_L} shows the comparison of the minimum jet power with $L_{\rm Edd}$ (top panel) and $L_{d}$ (bottom panel). None of the ISP and HSP sources in our sample have BH masses, thus their $L_{\rm Edd}$ is computed using the population estimates (Section~\ref{sec:method}). Except for a handful of sources with $P_{j,\min}\sim L_{\rm Edd}$ (within uncertainties), we find that the majority of blazars in the PS scenario has super-Eddington jet powers and  $P_{j,\min}\sim 10^2 L_d$ \citep[see also][]{Zdziarski2015}.

Figure~\ref{plt:P_v_P} (top panel) shows the comparison of $P_{j,\min}$ with $P_{\rm BZ}$ for 40 blazars with core-shift measurements (open colored symbols), assuming that all sources host maximally spinning BHs. We have also estimated the BZ power for sources without core-shift measurements (filled grey symbols) using the sample's median (and standard deviation) opening angle and magnetic field. Most sources cluster around the $P_{j,\min}=10^2 P_{\rm BZ}$ line, and the deviation from the line of equality becomes even larger when considering uniform or beta distributions for the BH spin (not shown in the figure). Meanwhile, we find that $\Phi_{\rm BH} \approx 50 \left(\dot{M} r_g^2 c\right)^{1/2} \propto L^{1/2}_d M_{\odot}$ (bottom panel), as expected for magnetically arrested accretion disks \citep{Bisnovatyi1974, Narayan2003}, in agreement with \cite{Zamaninasab2014}. Thus, the PS scenario predicts much higher jet powers than the BZ power, even in the MAD regime where the jet production efficiency is highest \citep{Tchekhovskoy2011}. While measured $P_{\rm BZ}$ are only available for 40 sources in our sample, the on-average estimates of the remaining sources follow the same trend well. Equation~(1) from \citealp{Zamaninasab2014} used to estimate $\Phi_{\rm jet}$, assumes energy equipartition between magnetic fields and radiating particles and does not explicitly consider the relation between the jet opening angle and magnetization $\sigma_M$. By relaxing this assumption, \cite{Zdziarski2015-II} derived a more general expression (see their Equation~21), which is identical to that of \citealp{Zamaninasab2014} for $\sigma_M=1$, but yields lower $\Phi_{\rm jet}$ values (by a factor of $2^{-1/2}$) for $\sigma_M \ll 1$. While a small correction given the uncertainty of individual estimates, it would only increase the discrepancy between $P_{\rm BZ}$ and $P_{j,\min}$.

Because of several assumptions made in this work (i.e., Doppler factor estimates, monoenergetic particle distribution, and proton radiative efficiency), the derived values of $P_{j,\min}$ constitute lower limits of the true minimum jet power further increasing this discrepancy. Hence, our results strongly disfavor the PS scenario for the majority of blazars, particularly for LSPs.

\section{Discussion}\label{sec:discuss}

\textit{Location of $\gamma$-ray emission region.} 
We can estimate the location of the $\gamma$-ray production site for the sources having estimates of the pc-scale jet's magnetic field as follows. Assuming that the jet's magnetic field is roughly equal to the magnetic field strength of the emission region, i.e., $B^\prime \approx B^\prime_{j,\phi} \propto 1/z$, we may write $z_{\rm em} \approx (B^\prime_{1 \rm pc}/B^\prime)~{\rm pc}$. We then find that $z_{\rm em}\ll 1$~pc, with 68\% of the values ranging between 0.006~pc and 0.08~pc, with a median of $0.03$~pc. Given that the median radius of the broad line region (BLR) for the sources of our sample is $0.15$~pc, our results suggest that the $\gamma$-ray production site should be well within the BLR.  This conclusion is, however, in tension with the lack of strong absorption features in the GeV $\gamma$-ray spectrum of luminous quasars \citep[e.g.,][]{Costamante2018}. The sub-pc location of the emission region is also inconsistent with the radius inferred by the average observed  variability, i.e.,  $r^\prime = c\dop \tvar/(1+z)$. The cross-sectional radius of the jet at the emission region can be written as $\varpi_{\rm em} \approx z_{\rm em} \theta_j$, for a conical jet with small half-opening  angle $\theta_j$ (the same assumption is made when computing $B^\prime_{1 \rm pc}$). Although a consistent picture would require $r^\prime  \lesssim \varpi_{\rm em}$, we find $r^\prime/\varpi_{\rm em}>1$, with 68\% of the ratio values in the range $9-60$ and a median of $27$. 

Part of this tension can be resolved, if one assumes that the magnetic field in the emission region is amplified with respect to the jet's toroidal magnetic field component. By writing $B^\prime = f_{\rm amp} B^\prime_{j,\phi}$ and requiring $r^\prime=\varpi_{\rm em}$, we find that the median amplification factor needed is 27. Thus, the $\gamma$-ray production site is also moved to pc scales, typically beyond the BLR (median $z_{\rm em}=0.5$~pc and 68\% of values ranging between 0.2~pc and 1.6 pc). Alternatively, lower $B^\prime$ values can be derived if the emission region moves with larger $\dop$ than what we have assumed (e.g., a three times higher $\dop$ for all sources would yield $B^\prime \sim 1-100$~G), but at the cost of higher $P_{j,\min}$.

\begin{figure}
\includegraphics[width=0.48\textwidth]{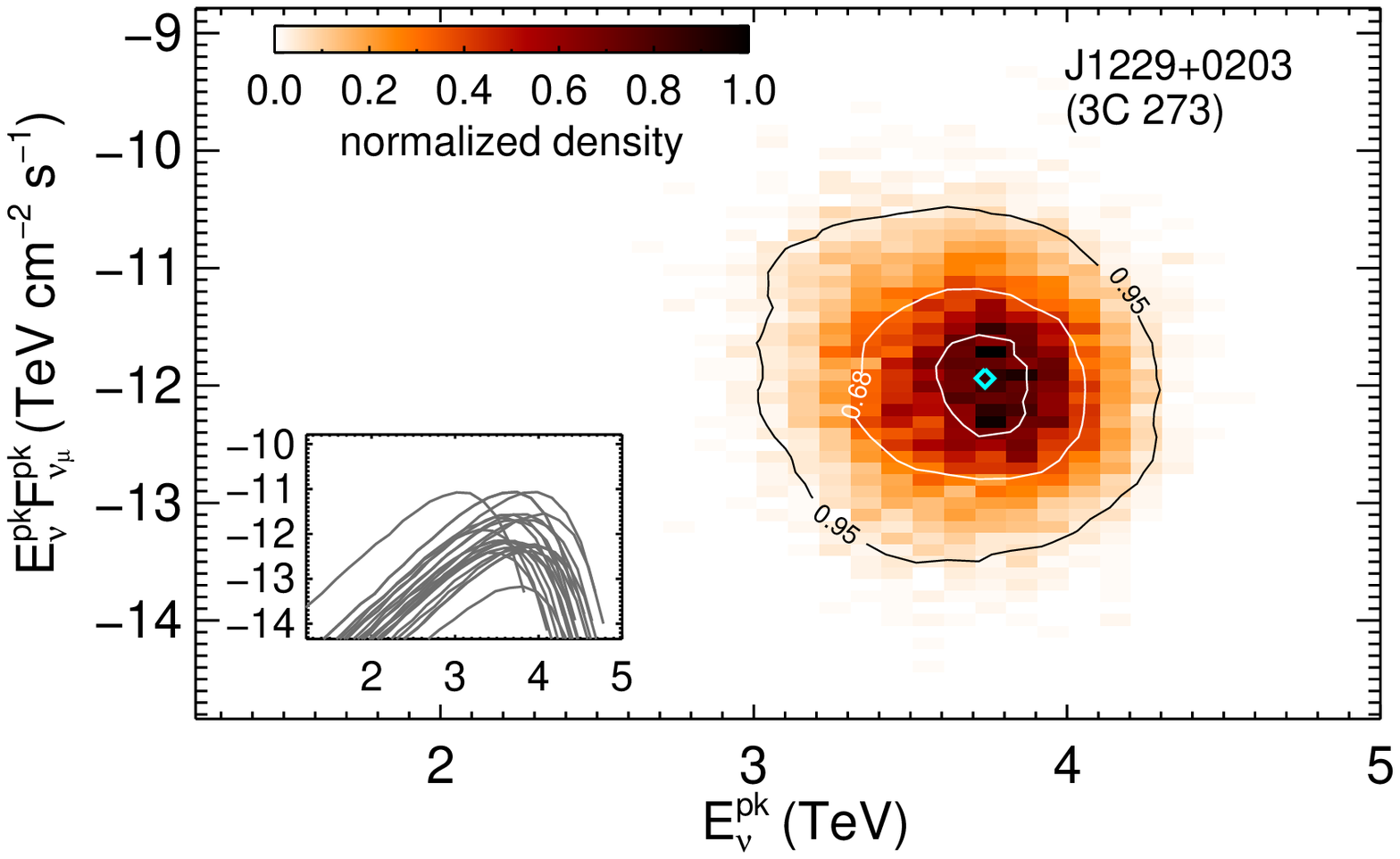}
\includegraphics[width=0.48\textwidth]{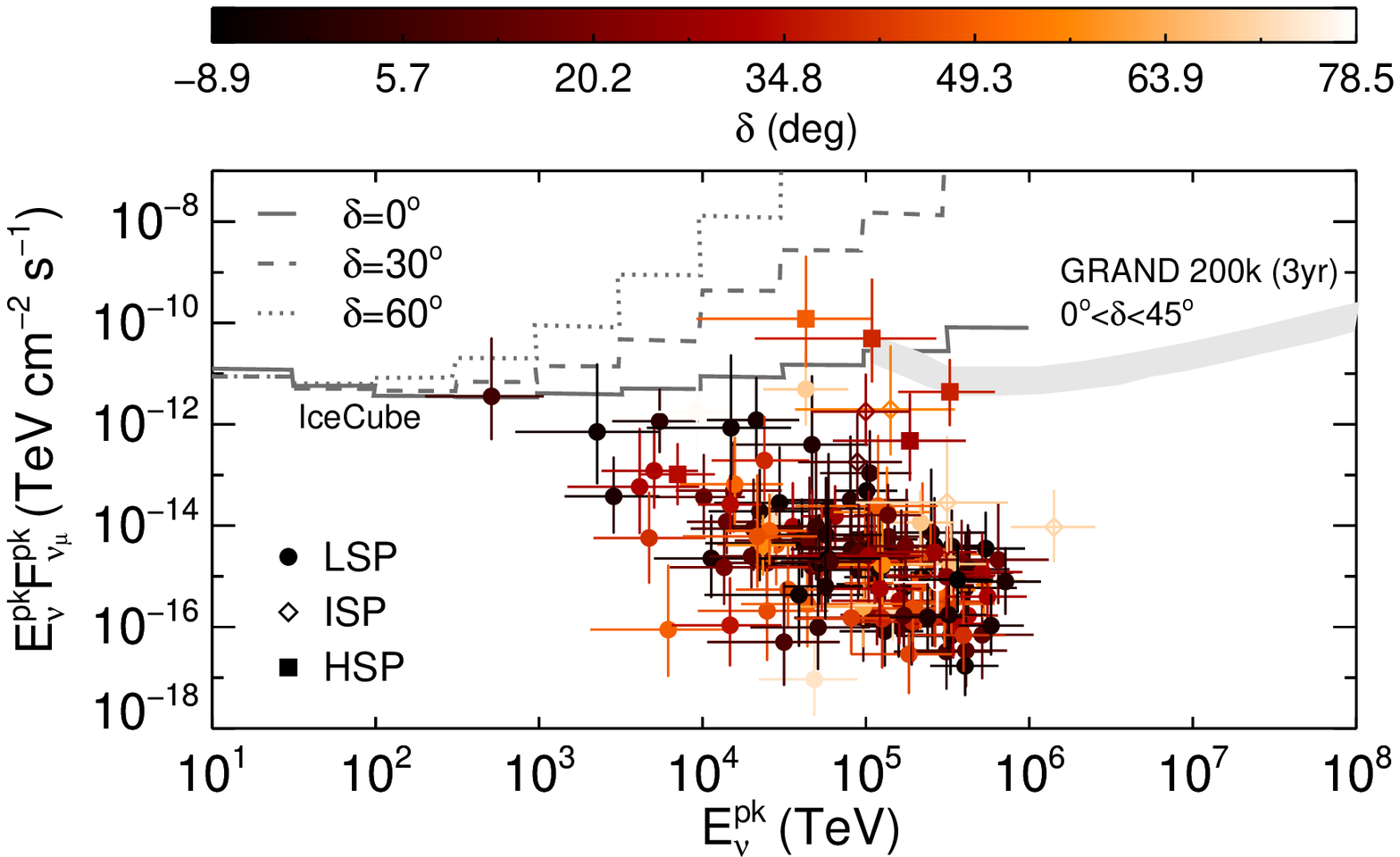}
 \caption{{\bf Top panel}: Density map of the predicted peak muon neutrino and anti-neutrino ($\nu_\mu + \bar{\nu}_{\mu}$) energy flux and peak neutrino energy for 3C~273 (inset panel shows individual neutrino energy spectra). The position of the median neutrino flux and energy is marked with an open diamond. The 68\% and 95\% density contours are also shown. {\bf Bottom panel}: Median peak $\nu_\mu + \bar{\nu}_{\mu}$ energy flux versus median peak neutrino energy for all the sources in our sample. The sources are color-coded according to declination ($\delta$) and the solid, dashed, and dotted lines show the IceCube 5$\sigma$ discovery potential for $\rm \delta=0^o$, $\rm \delta=30^o$, and  $\rm \delta=60^o$, respectively \citep{Aartsen2019EPJC}. The GRAND200k declination-averaged sensitivity to $\nu_{\mu}+\bar{\nu}_\mu$ for a 3-year observation window is also shown for comparison (gray colored band); adapted from  \cite{Grand2018}.}
\label{plt:neutrinos}
\end{figure}

\textit{High-energy neutrino emission.} 
Relativistic protons can also interact with low-energy photons to produce high-energy electron and muon neutrinos through the photo-meson ($\pg$) production process. The apparent isotropic proton luminosity $L_p$ and the absolute jet power in relativistic protons $P_{j,p}$ are related as $L_p = 2\dop^2 \psi^{-2} P_{j,p}$ \citep[e.g.,][]{dermer_2012}. For the purposes of this discussion, we replace the mononenergetic proton distribution with a power-law spectrum with an exponential cutoff, so that the differential apparent isotropic proton luminosity is written as,  $L_p(\epsilon_p) \propto \epsilon_p^{-p} e^{-\epsilon_p/\epsilon_{p,\max}}$, where $p=1.7$ and $\epsilon_{p,\max} = \dop \bar{\gamma}^\prime_p m_p c^2/(1+z)$. For every source, we compute $L_p(\epsilon_p)$ and $\epsilon_{p,\max}$ for parameters minimizing the total jet power  (Sections~\ref{sec:method} and \ref{sec:res}).  

Following our previous discussion on the location of the emission region, we assume that protons interact only with the jet's synchrotron photons. The differential number density of the low-energy photons is $n^\prime(x)=n_0^\prime \varepsilon_l^{\prime -2} \left[x^{-2+\Gamma_1}H(1-x) + x^{-2+\Gamma_2} H(x-1)\right]$,  where $\Gamma_1=1/2$, $\Gamma_2=-1/2$\footnote{We adopt the same photon indices for all sources, as a detailed calculation of the neutrino spectral shape lies beyond the scope of this work.}, $x\equiv\varepsilon^\prime/\varepsilon_l^\prime$, $\varepsilon^\prime_l = \varepsilon_l (1+z)/\dop$, and  $n_0^\prime = 3 L_l (1+z)^2 /4 \pi m_e c^5  \tvar^2 \dop^6$.  The $\pg$ efficiency is defined as $f_{\pg}\equiv\dop\tvar/t^\prime_{\pg}(1+z)$, where $t^\prime_{\pg}$ is the  energy-loss timescale. This is $t_{\pg}^{\prime -1}\left(\gamma^\prime_p \right)=c/(2\gamma_{p}^{\prime 2}) 
\int_{0}^{\infty} d\varepsilon' n'(\varepsilon')/\varepsilon^{' 2} \int_{\varepsilon_{th}}^{2\gp \varepsilon^\prime}\!\!\! d\varepsilon_{r} \sigma_{\pg}(\varepsilon_{r}) \kappa_{\pg}(\varepsilon_{\rm r}) \varepsilon_{r}$~\citep{stecker68}, where $\varepsilon_{th} \approx 400$ is the threshold photon energy for production of a $\Delta^+(1232)$ resonance,  $\kappa_{\pg}=0.2$ is the inelasticity  of interaction, and $\sigma_{\pg} \approx 0.34\, {\rm mb}$ for $\varepsilon_{th} \le \varepsilon_{r}\le 980$ is the cross section \citep{dermer_menon2009}.  The differential all-flavor neutrino (and anti-neutrino) flux is given by 
\eqb 
\epsilon_\nu F_{\nu+\bar{\nu}}(\epsilon_\nu) \approx \frac{3}{8} f_{\pg}\left(\frac{\epsilon_p (1+z)}{\dop m_p c^2}\right) \frac{\epsilon_p L_p(\epsilon_p)}{4\pi d_L^2}, 
\label{eq:Fnt}
\eqe 
where $\epsilon_\nu \approx \epsilon_p/20$ and $d_L$ is the luminosity distance. Because of neutrino oscillations the muon neutrino and anti-neutrino energy flux at Earth is $F_{\nu_\mu+\bar{\nu}_\mu}\approx F_{\nu+\bar{\nu}}/3$. 

Figure~\ref{plt:neutrinos} shows the peak neutrino energy and peak $\nu_\mu+\bar{\nu}_\mu$ energy flux derived from equation (\ref{eq:Fnt}). Our results are in line with predictions made for individual sources, i.e., $\sim0.1-1$ EeV neutrinos with fluxes much lower than in $\gamma$-rays \citep[e.g.,][]{Dimitrakoudis2014,Keivani2018}. There are a few blazars that are potentially interesting neutrino sources  (close to IceCube's discovery potential), with LSP blazar 4FGLJ2148.6+0652 being the best example.

For this blazar, we find  $B^\prime\sim2565$~G and $P_{j,\min} \gtrsim 10^2 L_d$, which is roughly three orders of magnitude higher than the average $P_{\rm BZ}$ value of LSPs with core-shift measuments. If steady neutrino emission at the predicted flux levels is detected from this blazar by IceCube, or from other sources by future experiments, such as GRAND \citep{Grand2018} or POEMMA \citep{Poemma} our understanding of accretion and jet launching in blazars needs to be revised.

\section{Conclusions}\label{sec:sum} 
We explored the energetic requirements for the proton synchrotron model for the largest sample of $\gamma$-ray blazars. The expectation for the minimum jet power in our sample far exceeds the observed $L_{\rm d}$ and $L_{\rm Edd}$ as well as the derived $P_{\rm  BZ}$, even more so, when considering that the results of this work constitute a lower limit to the true minimum jet power. In addition, the derived magnetic field strengths in the emission region imply either large  amplification of the jet's magnetic field or sub-pc $\gamma$-ray production sites, well within the BLR and in tension with recent results. The expected neutrino emission (for all sources in our sample) peaks at $\sim 0.1-10$~EeV, i.e., at much higher energies than the multi-TeV neutrinos associated with TXS 0506+056 \citep[see also][]{Keivani2018}. Meanwhile, the typical peak neutrino fluxes are $\sim 10^{-4}$ times lower than the peak $\gamma$-ray fluxes. Our results clearly demonstrate that the scenario where proton synchrotron accounts for the observed steady $\gamma$-ray emission in blazars is highly unlikely. Given that alternative hadronic models invoking emission from $\pg$ secondaries typically require even higher energy budgets \citep[e.g.,][]{Petropoulou2015}, we conclude that if a hadronic population is present in blazar jets it can only be a radiatively subdominant component or can dominate only during transient events.

\begin{deluxetable*}{llccccccccccc}
\tablenum{1}
\label{tab:par_sources}
\tablecaption{Parameter estimates for the sources in our sample.}
\tabletypesize{\scriptsize}%\scriptsize
\tablehead{\colhead{Name}& \colhead{Alt. name} & \colhead{$\epsilon_{h}$} & \colhead{$L_{h}$}& \colhead{$B^\prime$} & \colhead{$P_{j,\min}$ } & \colhead{$M_{\bullet}$}   & \colhead{$ L_{d}$}  & \colhead{$P_{\rm BZ}$}  & \colhead{$\epsilon^{pk}_{\nu}$}  & \colhead{$\epsilon^{pk}_{\nu}F^{pk}_{\nu_\mu+\bar{\nu}_{\mu}}$}}
\startdata
4FGLJ0017.5-0514 & J0017-0512 & 0.0332 & 45.05 & 16.88 $^{+14.75} _{-5.77}$ & 47.14 $^{+0.38} _{-0.21}$ & 7.55 $\pm$0.45 & 45.92 $\pm$0.51 & 45.86 $\pm$1.24 & 5.07 $^{+0.29} _{-0.19}$ & -15.15 $^{+1.64} _{-0.34}$  \\
4FGLJ0019.6+7327 & J0019+7327 & 0.0605 & 48.02 & 187.38 $^{+97.42} _{-64.10}$ & 48.42 $^{+0.38} _{-0.20}$ & 9.62 $\pm$0.52 & 47.15 $\pm$0.52 & 46.15 $\pm$1.21 & 4.43 $^{+0.24} _{-0.16}$ & -13.96 $^{+1.38} _{-0.32}$  \\
4FGLJ0051.1-0648 & J0051-0650 & 0.0937 & 47.27 & 168.76 $^{+311.88} _{-95.71}$ & 48.82 $^{+4.31} _{-0.38}$ & -- & -- & 46.86 $\pm$1.48 & 4.47 $^{+0.46} _{-0.26}$ & -13.55 $^{+5.07} _{-0.38}$  \\
4FGLJ0108.6+0134 & J0108+0135 & 0.1039 & 47.88 & 81.11 $^{+175.39} _{-46.00}$ & 49.26 $^{+2.39} _{-0.30}$ & -- & -- & 47.24 $\pm$1.36 & 4.77 $^{+0.56} _{-0.26}$ & -13.66 $^{+4.84} _{-0.39}$  \\
4FGLJ0112.8+3208 & J0112+3208 & 0.0147 & 46.44 & 59.26 $^{+64.03} _{-27.63}$ & 48.99 $^{+1.02} _{-0.24}$ & -- & -- & 46.94 $\pm$1.33 & 4.55 $^{+0.38} _{-0.23}$ & -14.02 $^{+2.74} _{-0.36}$  \\
4FGLJ0116.0-1136 & J0116-1136 & 0.0093 & 46.02 & 73.05 $^{+95.71} _{-34.07}$ & 48.94 $^{+1.09} _{-0.26}$ & 8.77 $\pm$0.38 & 45.32 $\pm$1.08 & 46.90 $\pm$1.36 & 4.35 $^{+0.39} _{-0.23}$ & -14.34 $^{+2.74} _{-0.36}$  \\
4FGLJ0132.7-1654 & J0132-1654 & 0.0178 & 46.75 & 53.37 $^{+57.67} _{-24.89}$ & 49.34 $^{+0.97} _{-0.27}$ & -- & -- & 47.36 $\pm$1.29 & 4.61 $^{+0.37} _{-0.21}$ & -14.28 $^{+2.34} _{-0.35}$  \\
4FGLJ0137.0+4751 & J0136+4751 & 0.0937 & 46.86 & 15.20 $^{+57.85} _{-9.27}$ & 48.40 $^{+0.77} _{-0.21}$ & 8.68 $\pm$0.31 & 46.23 $\pm$0.52 & 46.17 $\pm$1.07\textdagger & 5.35 $^{+0.63} _{-0.31}$ & -14.56 $^{+10.68} _{-0.39}$  \\
4FGLJ0152.2+2206 & J0152+2207 & 0.0314 & 46.49 & 187.38 $^{+293.26} _{-87.38}$ & 49.12 $^{+2.21} _{-0.30}$ & -- & -- & 46.98 $\pm$1.39 & 4.19 $^{+0.42} _{-0.24}$ & -13.31 $^{+3.54} _{-0.37}$  \\
4FGLJ0204.8+1513 & J0204+1514 & 0.0161 & 46.27 & 168.76 $^{+116.04} _{-57.73}$ & 49.36 $^{+0.60} _{-0.24}$ & -- & -- & 47.21 $\pm$1.24 & 4.15 $^{+0.28} _{-0.19}$ & -13.92 $^{+1.53} _{-0.33}$  \\
\enddata
\tablecomments{Table~1 is published in its entirety in machine-readable format. A portion is shown here for guidance regarding its form and content. The values of all parameters except $\epsilon_{h}$ and $B^\prime$ are displayed as $\log_{10}$. $\epsilon_{h}$ is in GeV, $\epsilon^{pk}_{\nu}$ in TeV,  $B^\prime$ in Gauss, $M_{\bullet}$ in solar masses, $\epsilon^{pk}_{\nu}F^{pk}_{\nu_\mu+\bar{\nu}_{\mu}}$ in TeV cm$^{-2}$ s$^{-1}$, and $L_{h}$, $P_{j,\min}$, $L_{d}$, $\rm P_{\rm BZ}$ in erg~s$^{-1}$.  All $P_{\rm BZ}$ estimates derived from core-shift measurements are indicated with \textdagger. 
}
\end{deluxetable*}

\acknowledgements
The authors would like to thank the anonymous referee for their useful comments and suggestions. The authors also thank A.~Mastichiadis,  E.~Resconi, and M.~Huber for useful discussions and comments. I.L. would also like to thank the Department of Astrophysical Sciences at Princeton University for its hospitality during which this project was conceived. This research has made use of data from the MOJAVE database that is maintained by the MOJAVE team  \citep{Lister2018}. M.P. acknowledges support from the Lyman Jr.~Spitzer Postdoctoral Fellowship and the  Fermi Guest Investigation grant 80NSSC18K1745. 

%\facilities{{\it Fermi}}

%\bibliographystyle{aasjournal}
% Use the LaTeX power, use bibtex properly.
%\bibliography{bibliography} %graphy.bib}%,bibliography_export.bib}

\end{document}